%% file: main.tex
\documentclass[conference]{IEEEtran}
\IEEEoverridecommandlockouts
\usepackage{cite}
\usepackage{amsmath,amssymb,amsfonts}
\usepackage{algorithmic}
\usepackage{graphicx}
\usepackage{textcomp}
\usepackage{xcolor}
\def\BibTeX{{\rm B\kern-.05em{\sc i\kern-.025em b}\kern-.08em
    T\kern-.1667em\lower.7ex\hbox{E}\kern-.125emX}}

\usepackage{amsmath,graphicx}
\usepackage{amsmath,amsfonts}
\usepackage{algorithmic}
\usepackage{algorithm}
\usepackage{array}
\usepackage{hyperref}
\usepackage[caption=false,font=normalsize,labelfont=sf,textfont=sf]{subfig}
\usepackage{textcomp}
\usepackage{stfloats}
\usepackage{url}
\usepackage{verbatim}
\usepackage{graphicx}
\usepackage{cite}
\usepackage{bm}
\usepackage{multirow}
\usepackage{amssymb}
\usepackage{comment}
\usepackage{mathtools}
\usepackage{booktabs}
\usepackage{balance}
\usepackage{flushend}
\usepackage{siunitx}
\usepackage{xcolor}
\usepackage{arydshln}
\AtBeginDocument{
  \abovedisplayskip     =1\abovedisplayskip
  \abovedisplayshortskip=1\abovedisplayshortskip
  \belowdisplayskip     =1\belowdisplayskip
  \belowdisplayshortskip=1\belowdisplayshortskip
}

\def\C{{\mathbb C}}
\def\R{{\mathbb R}}

\makeatletter
\def\bstctlcite{\@ifnextchar[{\@bstctlcite}{\@bstctlcite[@auxout]}}
\def\@bstctlcite[#1]#2{\@bsphack
  \@for\@citeb:=#2\do{%
    \edef\@citeb{\expandafter\@firstofone\@citeb}%
    \if@filesw\immediate\write\csname #1\endcsname{\string\citation{\@citeb}}\fi}%
  \@esphack}
\makeatother 

\let\OLDthebibliography\thebibliography
\renewcommand\thebibliography[1]{
  \OLDthebibliography{#1}
  \setlength{\parskip}{.5pt}
  \setlength{\itemsep}{1pt plus 0.3ex}
}

\begin{document}

\title{A Comparative Study on Positional Encoding for Time-frequency Domain Dual-path Transformer-based Source Separation Models}

\author{\IEEEauthorblockN{Kohei saijo\IEEEauthorrefmark{1},
Tetsuji Ogawa\IEEEauthorrefmark{1}}
\IEEEauthorblockA{\IEEEauthorrefmark{1}Department of Communications and Computer Engineering, Waseda University, Tokyo, Japan}
}

\maketitle

\begin{abstract}
In this study, we investigate the impact of positional encoding (PE) on source separation performance and the generalization ability to long sequences (length extrapolation) in Transformer-based time-frequency (TF) domain dual-path models.
The length extrapolation capability in TF-domain dual-path models is a crucial factor, as it affects not only their performance on long-duration inputs but also their generalizability to signals with unseen sampling rates.
While PE is known to significantly impact length extrapolation, there has been limited research that explores the choice of PEs for TF-domain dual-path models from this perspective.
To address this gap, we compare various PE methods using a recent state-of-the-art model, TF-Locoformer, as the base architecture.
Our analysis yields the following key findings:
(i) When handling sequences that are the same length as or shorter than those seen during training, models with PEs achieve better performance.
(ii) However, models without PE exhibit superior length extrapolation. This trend is particularly pronounced when the model contains convolutional layers.
\end{abstract}


\section{Introduction}
\label{sec:intro}

With the advancements in neural networks, source separation has witnessed remarkable progress over the past decade.
Early approaches, such as time-frequency (TF) masking~\cite{dc, pit}, have evolved into time-domain separation networks~\cite{tasnet,convtasnet}, which have gained significant attention.
More recently, TF-domain dual-path modeling has demonstrated superior performance, particularly in challenging noisy reverberant conditions~\cite{tfpsnet, tfgridnet}.
Simultaneously, a variety of network architectures, such as convolutional networks~\cite{convtasnet}, recurrent neural networks~\cite{dprnn}, Transformers~\cite{sepformer}, and state-space models~\cite{dpmamba}, have been explored for source separation.
Among these, Transformer-based models exhibit unique characteristics that are either unattainable or yet to be fully validated in other architectures, such as prompting to specify the task~\cite{uses, tuss}.
Consequently, Transformer-based models have attracted significant research interest, and recent TF-domain dual-path models have achieved state-of-the-art performance~\cite{tflocoformer}.

However, Transformer-based models often exhibit limited generalizability when applied to sequences longer than those seen during training.
This \textit{length extrapolation} problem is critical in source separation, as models are frequently required to process input sequences with a longer length than those seen during training.
In the context of TF-domain dual-path models, length extrapolation ability also affects generalizability to signals with unseen sampling rates.
Since short-time Fourier transform (STFT) with a fixed-duration window and hop size produces STFT spectra with a constant TF resolution regardless of the sampling rate, TF-domain dual-path models can handle inputs with various sampling rates~\cite{uses}.
However, as the number of frequency bins in the STFT representation changes with the input sampling rate, the model’s ability to generalize to unseen sampling rates is inherently dependent on its length extrapolation capability.

In Transformer-based models, positional encoding (PE) plays a crucial role, and the length extrapolation ability is highly dependent on the choice of PE.
For example, the absolute PE (APE) introduced in the original Transformer is known to exhibit poor generalization to longer sequences.
To mitigate this limitation, various PE methods have been proposed.
One such approach is rotary positional encoding (RoPE)~\cite{rope}, which is widely adopted in large language models (LLMs).
RoPE enhances length extrapolation by embedding positional information exclusively in the key and query vectors.
Some relative positional encoding (RPE) methods inspired by this concept have been proposed, further enhancing length extrapolation performance~\cite{alibi, chi2022kerple}.

Meanwhile, prior studies in computer vision have investigated the use of convolutional layers as PE, leveraging the fact that convolutional layers can implicitly encode positional information through zero-padding~\cite{conv_posenc}.
Since convolution-based PEs extract position information from local neighborhoods alone, they may demonstrate stronger generalization to longer sequences than those seen during training~\cite{conditional_pe}.
Given that many high-performing speech processing models incorporate both self-attention and convolution~\cite{conformer, tflocoformer}, they may be able to utilize position information encoded by convolution to enhance length extrapolation.
However, this approach remains unexplored in the context of source separation.
Indeed, although previous research has compared PEs in speech enhancement, it was limited to APE and RPE, without considering other PEs, such as convolution or RoPE.

In this work, we investigate the impact of positional encoding (PE) in Transformer-based TF-domain dual-path source separation models.
To this end, we examine several PE methods from different categories within the framework of TF-Locoformer~\cite{tflocoformer}, a state-of-the-art model.
TF-Locoformer incorporates both self-attention and convolution, enabling us to analyze not only the effects of explicit PEs, such as RPE, but also the implicit positional encoding induced by convolutional layers.
We systematically compare four types of PEs, absolute, relative, rotary, and convolution, and evaluate their impact on source separation performance and length extrapolation.

\section{Background}
\label{sec:background}

\subsection{TF-Locoformer}
\label{ssec:tflocoformer}

TF-Locoformer is based on TF-domain dual-path modeling, where frequency and temporal modeling are done alternately~\cite{tfpsnet, tfgridnet}.
Let us denote a $M$-channel mixture of $N$ sources in the STFT domain $\bm{S}\in \R^{2 \times N \times M \times T \times F}$ as $\bm{X} = \sum\nolimits_{n} \bm{S}_{n} \in \R^{2 \times M \times T \times F}$, where $T$ and $F$ are the number of frames and frequency bins, and $n=1,\dots,N$ is the source index.
2 corresponds to the real and imaginary (RI) parts.

The input $\bm{X}$ is first reshaped into ${2M \times T \times F}$ and then encoded into an initial feature $\bm{Z}$ with feature dimension $D$:
\begin{align}
   \label{eq:encode}
   \bm{Z} = \mathrm{gLN}(\mathrm{Conv2D}(\bm{X})) \in \R^{D \times T \times F},
\end{align}
where gLN is global layer normalization~\cite{convtasnet}.
For frequency modeling, we interpret the feature $\bm{Z}$ as a stack of T arrays, each with shape $D \times F$.
This is implemented by permuting the dimensions of $\bm{Z}$ to the form $T \times D \times F$.
Frequency modeling is then performed as follows:
\begin{align}
   \label{eq:sequence_modeling}
   \bm{Z} &\xleftarrow{} \bm{Z} + \mathrm{ConvSwiGLU}(\bm{Z}) / 2, \\
   \bm{Z} &\xleftarrow{} \bm{Z} + \mathrm{MHSA}(\mathrm{Norm}(\bm{Z})), \\
   \bm{Z} &\xleftarrow{} \bm{Z} + \mathrm{ConvSwiGLU}(\bm{Z}) / 2,
\end{align}
where $\mathrm{MHSA}$ stands for multi-head self-attention with $H$ heads~\cite{transformer}.
In the original TF-Locoformer, RoPE~\cite{rope} is used to encode the positional information of each frequency bin.
Root-mean-square group normalization (RMSGroupNorm), a derivative of RMSNorm~\cite{rmsnorm}, is adopted as the $\mathrm{Norm}$ layer.
The $\mathrm{ConvSwiGLU}$ module is a feed-forward network based on the convolution layers with a kernel size of $K$ and stride of $S$ and the swish gated linear unit (SwiGLU) activation~\cite{swiglu}: 
\begin{align}
   \label{eq:conv_swiglu}
   \bm{Z} &\xleftarrow{} \mathrm{Norm}(\bm{Z}), \\
   \bm{Z} &\xleftarrow{} \mathrm{Swish}(\mathrm{Conv1D}(\bm{Z})) \otimes \mathrm{Conv1D}(\bm{Z}), \label{eq:swiglu}\\
   \bm{Z} &\xleftarrow{} \mathrm{Deconv1D}(\bm{Z}).
\end{align}

Temporal modeling is done in the same way by viewing the feature $\bm{Z}$  as a stack of $F$ arrays of shape $D \times T$, where $T$ is the sequence length (i.e., we permute the dimension order of $\bm{Z}$ to $F \times D \times T$).
After alternating between frequency and temporal modeling $B$ times, the final feature $\bm{Z}$ is used to estimate the RI components of $N$ sources:
\begin{align}
   \label{eq:decode}
   \hat{\bm{S}} = \mathrm{DeConv2D}(\bm{Z}) \in \R^{2 \times N \times M \times T \times F}.
\end{align}

\subsection{Sampling-frequency independence (SFI) of TF-domain dual-path models}
\label{ssec:sfi}

In general, source separation models may struggle when processing input signals sampled at frequencies different from those used during training.
This performance degradation is primarily attributed to differences in input resolution.
However, when using STFT, the resolution of each TF bin remains unchanged as long as the fixed-duration STFT window and hop sizes are maintained, regardless of the input sampling frequency.
The STFT spectra of signals with different sampling rates but the same duration contain the same number of frames but a different number of frequency bins, while preserving a consistent resolution.
Leveraging this property, a separation model that does not explicitly depend on the number of frequency bins can exhibit sampling frequency invariance (SFI).
SFI allows, for example, a model trained at a lower sampling rate to be effectively applied to data with a higher sampling rate, thereby improving training efficiency.

Many TF-domain dual-path models inherently satisfy this requirement, as they employ sequence modeling that is independent of sequence length in the frequency dimensions.
Indeed, several models have been empirically shown to generalize well to sampling rates higher than those used during training~\cite{uses}.
TF-Locoformer also meets this criterion and is expected to exhibit SFI; however, its length extrapolation ability may strongly depend on the choice of PE.
In this study, we conduct a case study on TF-Locoformer, systematically comparing different PE methods in Transformer-based TF-domain dual-path models and evaluating their effectiveness in improving length extrapolation performance.

\section{Positional encodings}
\label{ssec:positional_encodings}

In this work, we investigate the following four PEs.

\subsection{Absolute positional encoding (APE)}
\label{ssec:ape}
As the APE, we use the sinusoidal positional encoding used in the original Transformer~\cite{transformer}.
The position embedding $\bm{P}_i \in \R^{D}$ at the position $i$ is defined as
\[
P_{d, i} =
\begin{cases}
\sin \left( 5000^{-\frac{d}{D}} \cdot i \right), & \text{if } d \text{ is even} \\
\cos \left( 5000^{-\frac{d-1}{D}} \cdot i \right), & \text{if } d \text{ is odd},
\end{cases}
\]
where $d=1,\dots,D$.
$\bm{P}$ is added to the encoded feature $\bm{Z}$ after the processing in Eq.~(\ref{eq:encode}).
As the TF-Locoformer has sequence modeling for both time and frequency dimensions, we add two positional embeddings separately. 
We first make a positional embedding for time frames $\bm{P}^{\mathrm{time}} \in \R^{D \times T}$, broadcast it to $D \times T \times F$, and add it to $\bm{Z}$.
We then make another embedding $\bm{P}^{\mathrm{freq}} \in \R^{D \times F}$, broadcast it to $D \times T \times F$, and add it to $\bm{Z}$.

\subsection{Kernelized relative positional embedding (KERPLE)}
\label{ssec:kerple}
Relative positional encoding (RPE) encodes the distances between each pair of elements in the input sequence.
It is known to generalize to longer sequences better than the APE.
Although there are many RPEs, we choose KERPLE~\cite{chi2022kerple}, which has shown better length generalization ability in speech enhancement\footnote{Although KERPLE is designed for causal processing, we decided to use it since it has been reported to work in non-causal speech enhancement~\cite{kerple_length_generalization_se}}~\cite{kerple_length_generalization_se}.
KERPLE kernelizes the distance between $i$-th and $j$-th elements in a sequence with conditionally positive definite kernels:
\begin{align}
   \label{eq:kerple}
   \bm{P}_{i,j} = -r_1(\mathrm{log}(1 + r_2|i-j|)),
\end{align}
where $r_1>0$ and $r_2>0$ are learnable parameters.
Unlike the APE, RPE is calculated in each MHSA layer, and $\bm{P}$ is added to the attention matrix.

\subsection{Rotary positional encoding (RoPE)}
\label{ssec:rope}
RoPE is used in the original TF-Locoformer to encode the position information.
RoPE encodes the position by applying rotation matrices to the key and query separately before computing the attention matrix.
Although RoPE is one of the most widely used PE and is employed in some LLMs~\cite{llama2}, recent works have shown that RoPE may not generalize to longer sequences than those seen during training~\cite{nope}.

\subsection{No positional encoding (NoPE)}
\label{ssec:nope}
In prior work in the field of computer vision, it has been shown that convolutional layers inherently encode positional information~\cite{conv_posenc}, allowing Transformers with convolutional layers to function effectively even without explicit PE~\cite{wang2022pvt}.
Removing PE not only reduces computational costs but also may facilitate generalization to sequences longer than those encountered during training~\cite{conditional_pe}.
Inspired by these findings, this study investigates TF-Locoformer with no positional encoding (NoPE).
Since TF-Locoformer incorporates convolutional layers, it is likely to work even in the absence of PE.
We investigate whether the model generalizes better to long sequences compared to its counterpart that includes PE.

\section{Experiments}
\label{sec:experiments}

\subsection{Datasets}
\label{ssec:datasets}

\textbf{WHAMR!}~\cite{whamr} contains noisy reverberant two-speaker mixtures sampled at 8 kHz.
Only the first channel was used to evaluate performance in a monaural setup.
Speech and noise signals are sourced from WSJ0~\cite{wsj0} and WHAM!~\cite{wham}, respectively.
The dataset comprises 20000, 5000, and 3000 mixtures for training, validation, and testing, respectively.
The model is trained to perform dereverberation, denoising, and separation simultaneously.

\noindent\textbf{MUSDB18-HQ} contains 100 and 50 stereo songs sampled at 44.1 kHz for training and testing, respectively.
For data partitioning, we adopted a commonly used split, allocating 86 songs for training and 14 for validation.
The goal is to separate the mixture into the individual stems (vocals, bass, drums, and other).
The model takes stereo mixtures as input and estimates stereo-separated signals.

\subsection{Model configuration}
\label{ssec:model_conf}

When training on WHAMR!, we used the small version of the TF-Locoformer~\cite{tflocoformer}.
Specifically, unless otherwise stated, we set the parameters as follows: $D=96$, $B=4$, $K=8$, $S=1$, and $H=4$ (notation follows Section~\ref{ssec:tflocoformer}).

For training on MUSDB, to reduce computational cost, we replace the Conv2D and DeConv2d layers in Eq.~(\ref{eq:encode}) and Eq.~(\ref{eq:decode}), respectively, with the band-split encoder and band-wise decoding module~\cite{bsrnn}.
The band-split encoder decomposes the input spectrogram $\bm{X} \in \R^{2M \times T \times F}$ with $F$ frequency bins into $Q$ non-overlapping subband spectrograms $\bm{X}_{q} \in \C^{T \times b_q}$ $(q={1,\dots ,Q})$, where the pre-defined band-widths $b_{q}$ satisfy $\sum_{q} b_{q} = F$.
$\bm{X}_{q}$ undergoes normalization and a linear transformation, producing a feature representation
$\bm{Z}_{q} \in \R^{D \times T \times 1}$.
The $Q$ features are then concatenated and result in a feature $\bm{Z} \in \R^{D \times T \times Q}$, which is processed by TF-Locoformer blocks.
The band-wise decoding module splits $\bm{Z}$ into $Q$ sub-features and decodes them to obtain band-wise masks (see \cite{bsrnn} for more details).
We used the same band-split configuration as~\cite{bsroformer} with $Q=62$ bands and refer to this model as the band-split Locoformer (BS-Locoformer).

\subsection{Training and evaluation details}
\label{ssec:train_eval_details}

The training was conducted with a single Nvidia RTX A5000 GPU.
For WHAMR!, we trained the TF-Locoformer for up to 150 epochs, with each epoch consisting of 5000 training steps.
We used the AdamW optimizer~\cite{adamw} with a weight decay factor of 1e-2.
The learning rate was linearly increased from 0 to 1e-3 over the first 4000 training steps, kept constant for the first 75 epochs, and subsequently decayed by a factor of 0.5 if the validation loss did not improve for three consecutive epochs.
Early stopping was applied when the best validation score remained unchanged for ten epochs.
The batch size was four.
Gradient clipping was applied with a maximum gradient $L_2$-norm of five.
The negative scale-invariant signal-to-distortion ratio (SI-SDR)~\cite{sisdr} was used as the loss function.
No data augmentation was applied.
To enhance training efficiency, we employed the automatic mixed precision technique and flash attention~\cite{dao2022flashattention}.

For MUSDB, we trained the BS-Locoformer for 200 epochs without early stopping, with each epoch consisting of 2500 training steps.
The optimizer and learning rate scheduling followed the WHAMR! configuration, except for the learning rate decay factor of 0.8.
We applied an unsupervised source activity detection method from~\cite{bsrnn} with a segment duration of eight seconds to remove silent segments from the training data.
During training, dynamic mixing was performed at each training step, where a segment from each stem was randomly selected, RMS-normalized, scaled by a gain uniformly sampled from [-10, 10] dB, and mixed.
Following \cite{bsrnn}, each segment was dropped with a probability of 0.1 before mixing to simulate the mixture where the target source is inactive.
The loss function was the negative SNR loss which accepts zero signals as ground truth~\cite{fuss}.
During inference, each song was segmented into multiple $T'$-second chunks with half overlap, separated individually, and reconstructed using overlap-add to obtain song-level separation results.
We use the utterance-level SDR (uSDR) as the evaluation metric~\cite{mdx2021}.
All other configurations remained consistent with those used for the WHAMR! setup.

\subsection{Results on WHAMR!}
\label{ssec:results_whamr}

\input{tables/whamr}
\input{tables/whamr_ks1}
\input{tables/whamr_sota}

Table~\ref{table:whamr} presents the average SI-SDR improvement (SI-SDRi) on the WHAMR! test set.
Both \texttt{A*} and \texttt{B*} used the 8kHz min version of training data; however, \texttt{A*} models were trained with one-second input segments, whereas \texttt{B*} models were trained with four-second input segments.

As shown in Table~\ref{table:whamr}, the NoPE approach achieves performance comparable to or better than other explicit PE methods under matched conditions (8k min).
In contrast, Table~\ref{table:whamr_ks1} shows the results for models without convolution, where the convolutional kernel size $K$ is set to one, demonstrating that some explicit PEs outperform NoPE on 8k min data.
These results suggest that in models incorporating convolutional layers, such as TF-Locoformer, convolution implicitly encodes position information, allowing explicit PE to be omitted without compromising overall performance.
Notably, in our experimental setup, training \texttt{B4} (NoPE) was approximately 16\% faster than training \texttt{B3} (RoPE).
To further assess whether NoPE maintains its effectiveness in larger models, we evaluated a medium-sized TF-Locoformer, with results summarized in Table~\ref{table:whamr_sota}.
The NoPE-based model outperformed the RoPE-based model and other state-of-the-art models, demonstrating its scalability.

Next, we analyzed the length extrapolation capability of time-frame modeling by focusing on the results of \texttt{A*}.
A comparison between the `min` and `max` data reveals that KERPLE and NoPE achieve the best extrapolation performance.
The strong extrapolation ability of KERPLE aligns with findings from previous studies~\cite{kerple_length_generalization_se}; however, our results also confirm that NoPE, when used in conjunction with convolutional layers, is equally effective.

We then examine the length extrapolation capability in frequency modeling by comparing results on the 8kHz and 16kHz test sets.
Similarly, NoPE and KERPLE exhibit the best performance, with NoPE achieving slightly superior results.
Compared to RoPE (\texttt{B3}), which was originally used in TF-Locoformer, NoPE (\texttt{B4}) performs comparably on the 8kHz min dataset while improving SI-SDR by approximately 0.4 dB on the 16kHz min dataset.
These findings suggest that convolution-encoded positional information also enhances frequency extrapolation.

Although NoPE demonstrates the best length extrapolation capability, an interesting trend emerges in Table~\ref{table:whamr}: for input sequences shorter than those used during training, explicit PE methods outperform NoPE.
This effect is evident in the performance of \texttt{C1} and \texttt{C2}, which were trained on 16 kHz data and evaluated on 8 kHz data, where \texttt{C1} achieves superior results.
Based on these findings, we conclude that NoPE is preferable for handling sequences longer than those seen during training, whereas explicit PE methods, such as RoPE, are more suitable for shorter sequences.

\subsection{Results on MUSDB}
\label{ssec:results_mss}

\begin{figure}[t]
\centering
\vspace{0mm}
\centerline{\includegraphics[width=0.9\linewidth]{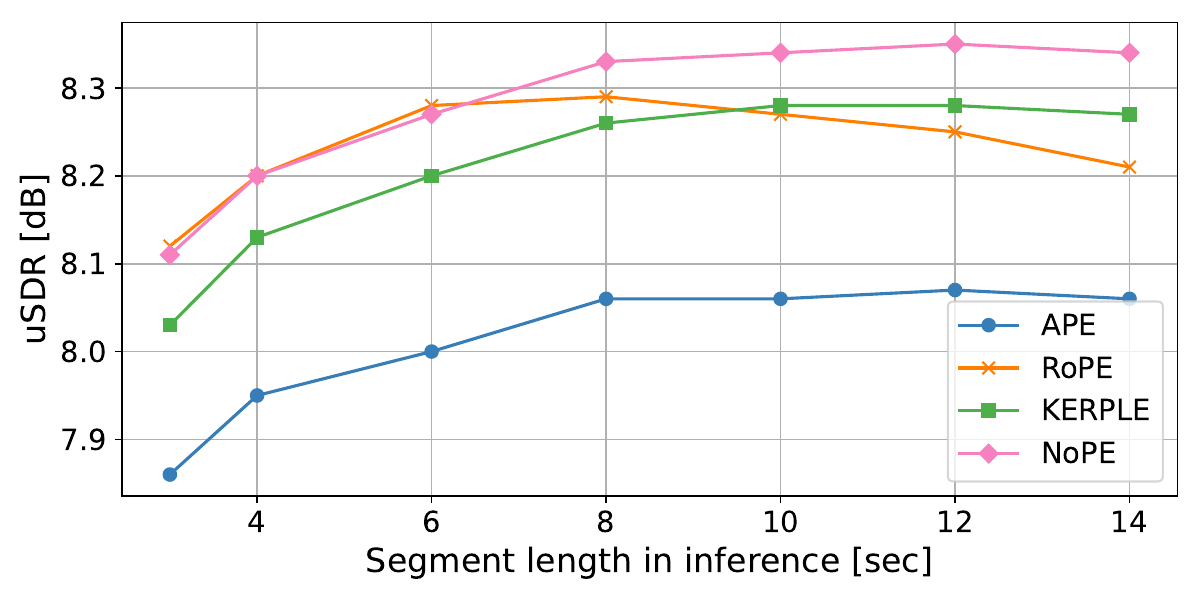}}
\vspace{-4mm}
\caption{
   uSDR scores of models trained on 3-second chunks when changing segment length in inference.
   Hop size was set to half of segment length.
}
\label{fig:mss_sdr_train3s}
\vspace{-5mm}
\end{figure}

\begin{figure}[t]
\centering
\centerline{\includegraphics[width=0.9\linewidth]{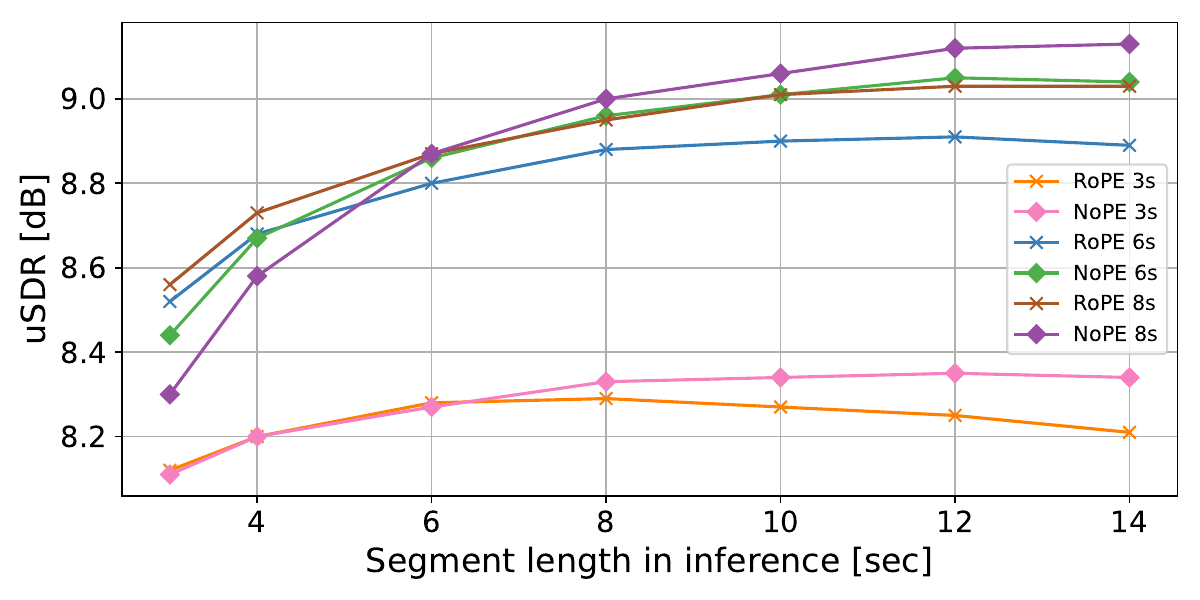}}
\vspace{-4mm}
\caption{
   uSDR scores of models trained on 3-, 6-, or 8-second chunks.
   uSDRs are evaluated changing segment length in inference, where hop size was set to half of segment length.
}
\label{fig:mss_sdr_train3-6-8s}
\vspace{-5mm}
\end{figure}

Figure \ref{fig:mss_sdr_train3s} presents the uSDR scores of BS-Locoformer trained with three-second segments.
The horizontal axis represents the segment length $T'$ used during inference, which was set to 3, 4, 6, 8, 10, 12, and 14 seconds for evaluation.
As shown in the figure, when $T'=3$, matching the training condition, NoPE and RoPE achieve the best performance.
As $T'$ increases, performance improves; however, while RoPE saturates at $T'=8$, NoPE continues to improve beyond $T'>8$.
Although KERPLE exhibits lower overall performance, it follows a similar trend to NoPE, further demonstrating the strong length extrapolation capabilities of NoPE and KERPLE.

Figure \ref{fig:mss_sdr_train3-6-8s} presents the evaluation results for models trained with segment lengths of 3, 6, or 8 seconds, using either RoPE or NoPE.
Regardless of the PE method, models trained with longer input segments tend to achieve better performance.
However, even under these conditions, NoPE exhibits superior length extrapolation compared to RoPE.
In contrast, when $T'$ is shorter than the training segment length, RoPE outperforms NoPE.
This observation is consistent with findings from WHAMR! experiments, further suggesting that explicit PE methods such as RoPE are preferable for handling sequences shorter than those seen during training.

\section{Conclusion}
\label{sec:conclusion}

In this work, we investigated the impact of positional encoding (PE) on length extrapolation in Transformer-based TF-domain dual-path models.
To this end, we compared four PE methods, APE, KERPLE, RoPE, and NoPE, in a framework of the TF-Locoformer, a recent state-of-the-art model.
Through experiments on both speech separation and MSS, we observed the following trends:
(i) When handling sequences that are the same length as or shorter than those seen during training, models with explicit PEs tend to perform better.
(ii) However, models without PE exhibit superior length extrapolation, a trend that is particularly pronounced when the model includes convolutional layers.

\section{Acknowledgement}
\label{sec:ack}
This work was supported by JSPS KAKENHI Grant Number JP24KJ2096.

\bibliographystyle{IEEEtran}
\bibliography{main}

\end{document}

%% file: tables/whamr.tex
\begin{table}[t]
\sisetup{
detect-weight, %
mode=text, %
tight-spacing=true,
round-mode=places,
round-precision=1,
table-format=2.1,
table-number-alignment=center
}
\begin{center}
\vspace{0mm}
\caption{
    Average SI-SDRi on WHAMR! test set.
    `min` version is used for training, while evaluation is done on both `min` and `max` versions, each with 8kHz and 16kHz data.
}
\vspace{-5mm}
\label{table:whamr}
\resizebox{\linewidth}{!}{
\begin{tabular}{lcclS[table-format=3.1]SSSS}
\toprule

& {Train} &{Input} & & \multicolumn{2}{c}{8kHz} & \multicolumn{2}{c}{16kHz} \\
ID &{SF} & {len[s]} &{PE} &{min} &{max} &{min} &{max}  \\

\midrule

    \texttt{A1} &8kHz &1 &APE &15.85 &16.43 &6.44 &5.36 \\
    \texttt{A2} &8kHz &1 &KERPLE &16.10 &16.72 &15.31 &15.85 \\
    \texttt{A3} &8kHz &1 &RoPE   &15.83 &16.16 &14.89 &14.96 \\
    \texttt{A4} &8kHz &1 &NoPE   &16.05 &16.72 &15.42 &15.99 \\

\hdashline

    \texttt{B1} &8kHz &4 &APE &17.58 &18.40 &14.80 &14.79 \\
    \texttt{B2} &8kHz &4 &KERPLE &17.67 &18.44 &16.96 &17.68 \\
    \texttt{B3} &8kHz &4 &RoPE   &\bfseries 17.78 &18.53 &16.88 &17.63 \\
    \texttt{B4} &8kHz &4 &NoPE   &\bfseries 17.77 &\bfseries 18.64 &17.25 &18.14 \\

\midrule
    \texttt{C1} &16kHz &4 &RoPE &17.41 &18.18 &17.64 &18.51 \\
    \texttt{C2} &16kHz &4 &NoPE &17.32 &17.93 &\bfseries 17.73 &\bfseries 18.68 \\

\bottomrule

\end{tabular}
}
\end{center}
\vspace{-5mm}
\end{table}

%% file: tables/whamr_ks1.tex
\begin{table}[t]
\sisetup{
detect-weight, %
mode=text, %
tight-spacing=true,
round-mode=places,
round-precision=1,
table-format=2.1,
table-number-alignment=center
}
\begin{center}
\vspace{0mm}
\caption{
    Average SI-SDRi when $K=1$ on WHAMR! test set.
    Training is done with 8kHz `min` version, while testing is done using both `min` and `max` versions.
    Training chunk was 4-seconds.
}
\vspace{-2mm}
\label{table:whamr_ks1}
\resizebox{0.7\linewidth}{!}{
\begin{tabular}{llS[table-format=3.1]SSSS}
\toprule

& &\multicolumn{2}{c}{8kHz} &\multicolumn{2}{c}{16kHz} \\
ID &{PE} &{min} &{max} &{min} &{max}  \\

\midrule

    \texttt{D1} &APE &13.35	&12.57 &4.30 &1.51 \\
    \texttt{D2} &KERPLE &13.82 &13.66 &9.93 &9.79 \\
    \texttt{D3} &RoPE   &\bfseries 13.94 &\bfseries 14.36 &6.58 &6.65 \\
    \texttt{D4} &NoPE   &13.48 &\bfseries 14.36 &\bfseries 11.19 &\bfseries 11.97 \\

\bottomrule

\end{tabular}
}
\end{center}
\vspace{-5mm}
\end{table}

%% file: tables/whamr_sota.tex
\begin{table}[t]
\sisetup{
detect-weight, %
mode=text, %
tight-spacing=true,
round-mode=places,
round-precision=1,
table-format=2.1,
table-number-alignment=center
}
\begin{center}
\caption{
    Comparison with previous models on WHAMR!. 
    DM$^*$ indicates DM with speed perturbation. Results in [dB].
    $\dagger$ denotes that the result is reproduced by us.
}
\vspace{-3mm}
\label{table:whamr_sota}
\resizebox{0.65\linewidth}{!}{
\begin{tabular}{lcS[table-format=3.1]SS}
\toprule

{System} & {SI-SDRi} & {SDRi}  \\

\midrule

    SepFormer+DM$^*$~\cite{sepformer} &14.0 &13.0  \\

    MossFormer2+DM$^*$~\cite{mossformer2} &17.0 &{-}  \\ 

    TF-GridNet~\cite{tfgridnet} &17.1 &15.6  \\ 

    TF-Locoformer (S)~\cite{tflocoformer} &17.4 &15.9  \\

    TF-Locoformer (M)~\cite{tflocoformer} &18.5 &16.9  \\

\midrule

    TF-Locoformer (S)$^\dagger$ &17.8 &16.1  \\

    TF-Locoformer-NoPE (S) &17.8 &16.1  \\

    TF-Locoformer-NoPE (M) &\bfseries 18.8 &\bfseries 17.1  \\

\bottomrule

\end{tabular}
}
\end{center}
\vspace{-5mm}
\end{table}